\documentclass[aps, twocolumn,showpacs,preprintnumbers,amsmath]{revtex4}
\usepackage{dcolumn}
\usepackage{bm}
\usepackage{graphicx}
\usepackage{bm}
\begin{document}
\title{Field induced magnetic transition and metastability in Co substituted $Mn_{2}Sb$}
\author{Pallavi Kushwaha, R Rawat and P Chaddah}
\affiliation{UGC-DAE Consortium for Scientific Research\\University Campus, Khandwa Road\\
Indore-452017, India.}
\date{\today}

\begin{abstract}
A detailed investigation of first order ferrimagnetic (FRI) to antiferromagnetic (AFM) transition in Co ($15\%$) doped $Mn_2Sb$ is carried out. These measurements demonstrate anomalous thermomagnetic irreversibility and glass-like frozen FRI phase at low temperatures. The irreversibility arising between the supercooling and superheating spinodals is distinguised in an ingenious way from the irreversibility arising due to kinetic arrest. Field annealing measurements shows reentrant FRI-AFM-FRI transition with increasing temperature.  These measurements also show that kinetic arrest band and supercooling band are anitcorrelated i.e regions which are kinetically arrested at higher temperature have lower supercooling temperature and vice versa.
\end{abstract}


\maketitle
\section {Introduction}
The compound $Mn_2Sb$ crystallizes in $Cu_2Sb$ type tetragonal structure and orders ferrimagnetically (FRI)  below $T_C\approx550 K$ \cite{beck, wilk}. In $Cu_2Sb$ type crystal structure there are two crystallographically non equivalent sites for Mn atoms, Mn(I) and Mn(II), which are tetrahedrally and octahedrally surrounded by Sb.  The magnetic structure of this compound consists of Mn(I)-Mn(II)-Mn(I) triple layer where magnetic moments of Mn(I)  ($(2.1 \mu _B)$and Mn(II) ($3.9 \mu _B$) are aligned antiparallel to each other.  These triple layer align parallel to each other resulting in the ferrimagnetic state. The substitution of various elements for Mn or Sb results in the appearance of first order FRI to antiferromagnetic (AFM) transition with decreasing temperature \cite{kano, swob, bith, darn, wijn, bart, zhan}. In the AFM state Mn(I)-Mn(II)-Mn(I) triple layers align antiparallel to each other \cite{aust}. The FRI to AFM transition is accompanied with large change in unit cell volume, resistivity, magnetization etc. \cite{bart, zhan, bar, zha, bier} The observed AFM to FRI transition in doped $Mn_2Sb$ based compound were interpreted in terms of exchange inversion model of Kittle \cite {kitt}. According to this model the exchange integral depends on the interatomic distance and changes sign for a critical value. However some recent studies have suggested itinerant nature of magnetism and role of electronic density of states in driving the transition. The electronic specific heat coefficient ($\gamma$) is found to increase in the FRI state as compared to AFM state \cite{bart, zha, baran} and magnetic moment on Mn(II) decreases by $(\approx 1\mu _B)$ across the AF to FRI transition \cite{ohas}. 

The first order AFM to FRI transition shows many interesting and sometime anomalous properties. Below FRI to AF transition temperature $T_N$, application of magnetic field induces a first order AF to FRI transition which give rise to large magnetoresistance, magnetostriction etc. \cite{zhan, bar, zha}. Due to large magnetic field effect and tunability of transition temperature up to and above room temperature these compounds have been of interest.  However, in spite of many studies on the doped $Mn_2Sb$ there are very few studies on the low temperature behavior of these systems. There are few reports which do indicate anomalous low temperature behavior in these systems. Zhang et al. (in $Mn_{2}Sb_{0.95}Sn_{0.05}$)  \cite{zha} and Bartashevich et al. (in $Mn_{1.80}Co_{0.20}Sb$) \cite{bart} have observed virgin curve lying outside  envelope curve in their magnetization measurements at low temperatures.  The H-T phase diagram based on high magnetic field studies in the Co substituted $Mn_2Sb$ shows a broadening of hysteretic region with decreasing temperature and a shallow maxima around 40 K for lower critical field (field required for FRI to AFM transition) \cite{bart}. Similar non-monotonic behavior for lower critical field (ferromagnetic to AFM transition) has been observed at low temperature for $Nd_{0.5}Sr_{0.5}MnO_3$ \cite{kuwa, toku, rawa} which shows an anomalous thermomagnetic irreversibiltiy. Such anomalous behavior in the first order transition at low temperature has been of current interest for its implications in the physics of manganites and glasses \cite {mane, bane, ban}.

In the present study we carried out detailed investigation of first order transition in $Mn_{1.85}Co_{0.15}Sb$ at low temperatures. This study shows that critically slow dynamics of phase transition at low temperature results in glass-like kinetic arrest \cite{mane, bane, ban} of high temperature FRI phase. The observed irreversibility at low temperatures (due to kinetic arrest) and irreversiblity around transition temperature (due to supercooling and superheating effect) are distinguished. By following novel paths in the H-T space one can observe a glass-like FRI phase at low temperature and the system shows a reentrant FRI to AFM to FRI transitions with increasing temperature. These measurements also shows tunability of coexisting FRI and AFM phase fractions and reveal an anti-correlation between supercooling and kinetic arrest bands. This system is the first FRI system which can be placed along side with intermetallic systems Al doped $CeFe_2$ \cite{mane, sing}, $Gd_5Ge_4$ \cite{chat}, $Nd_7Rh_3$ \cite{sen} where such an arrest of kinetics was observed.

\maketitle\section{Experimental Details}

The compound $Mn_{1-x}Co_xSb$ with $x=0.15$ is prepared by arc melting the constituent elements of purity better than $99.99\%$ (LIECO Industries, USA) under high purity Argon atmosphere. Powder X-ray diffraction of the prepared compound shows that compound crystallizes in $Cu_2Sb$ type tetragonal structure and there are no impurity peaks. The resistivity measurements are performed by standard four-probe technique using a commercial cryostat (Oxford Instruments Inc., UK) in the temperature range $3-300$ K and up to $8$ tesla magnetic fields. For in-field measurements magnetic field is applied parallel to the current direction.

\maketitle\section{Results and discussion}
Figure 1 shows the temperature dependence of resistivity in the absence of magnetic field for $Mn_{1.85}Co_{0.15}Sb$. A sharp rise in resistivity with decreasing temperature indicates transition from low resistance FRI to high resistance AFM phase. It must be noted here that both the phases are metallic in nature and negative resistivity slope in the transition region is due to increasing AFM phase fraction (high resistivity) with lowering temperature. During warming, transition occurs at higher temperature resulting in a thermal hysteresis across the transition. It indicates the first order nature of the transition. Besides this, it is also observed that room temperature resistivity increases ($\approx 1 \%$)on the initial thermal cycling through the transition. It is well known that in this compound the AFM to FRI transition is accompanied with large volume change which can result in microcracks and hence increased resistivity during the transition \cite{bart}. We have taken care of it during interpretation of our result and such effects are minimized by repeated thermal cycling of the sample. The derivative of temperature dependence of resistivity is shown as an inset of figure 1 which shows a sharper transition during heating compared to during cooling cycle. Similar asymmetry has been observed ealrier in $CeFe_{0.96}Ru_{0.04}$ \cite{chatto} and $FeRh$ \cite{maat} and attributed to nucleation and growth process during a first order phase transition. The transition temperature which is taken as the inflection point of the resistivity curve is found to be $\approx 118 K$ during cooling and $\approx 132 K$ during warming cycle and these are in good agreement with reported transition temperature for this composition \cite{bart, bar}. As can be seen from figure 1 the transition width is broad ($\approx 60 K$) during both cooling and heating. The broadening of first order transition can arise due to chemical inhomogeneity or disorder inherent in substitutional alloys. In a disorder-free sample the first order transition will occur at a sharply defined $(H_C, T_C)$ line. Due to disorders different regions, having length scale of the order of the correlation length can have different transition temperatures, and this results in transition line broadening into a band \cite{hern, cha, imry}. Supercooling and superheating spinodals will also form a band for such a sample \cite{cha}. Though broad transitions are undesirable for many application but in present study it will be useful in studying phase coexistence and correlation between supercooling and kinetic arrest band. It is illustrated in figure 1 (b), which shows schematic diagram of kinetic arrest band and supercooling band adapted from Banerjee et al. \cite{ban}. The interplay between the kinetic arrest and supercooling takes place when field cooling is done with fields lying between $H_1$ and $H_2$ resulting in a coexisting arrested (FRI) and stable (AFM) phase at low temperature \cite{ban}. If the band widths are narrow then the field window $(H_2 - H_1)$ will decrease and become zero in the limit of zero band width so that there will be no phase coexistence; system will be either in completely AFM state or in completely arrested FRI state. 
 
The FRI to AFM transition temperature is suppressed with the application of magnetic field which can be seen from figure 2 showing resistivity behavior in presence of various magnetic fields. This figure not only shows decreasing transition temperature with magnetic field but also highlights the history dependence of resistivity behavior. To study the path dependence of the resistivity, measurements were carried out under following sequence: First sample is cooled under zero field to lowest temperature of measurement and then labeled magnetic field is applied isothermally and (i) resistivity is measured during warming under applied constant magnetic field up to 300K indicated as ZFCW; then (ii) measured during cooling under same magnetic field (FCC); and subsequently (iii) measured during warming under same magnetic field (FCW). The resistivity behavior below $T_N$ in  figure 2 shows that ZFCW curves have higher resistivity as compared to FC curves. Here it will be worth mentioning that the higher resistivity for ZFCW curve compared to FC curve is intrinsic property of the sample. It is not a thermal cycling effect,  since ZFCW curve which has higher resistivity has been measured before FC curve. With increasing magnetic field the difference between ZFCW and FC curve increases. For 8 Tesla field there is no signature of FRI to AFM transition for FC measurement. Such thermomagnetic irreversibility for resistivity in ZFCW and FC curve has been observed in many other system across the first order transition like in transition metal doped $CeFe_2$ \cite{mane, sing}, $Nd_{0.5}Sr_{0.5}MnO_3$ \cite {rawa} etc. In these systems it has been attributed to coexisting phases whose ratio depends on the path traversed in the H-T space. 

The isothermal magnetoresistance (R-H) at various temperatures below $T_N$ are shown in figure 3. For these measurements samples were cooled under zero field from well above $T_N$ to the measurement temperature. With the application of magnetic field a field induced transition from AFM to FRI phase is observed as a sharp decrease in resistivity. The field at which AFM to FRI transition is observed increases with decreasing temperature. In fact, for temperature $T\leq 50K$, 8 Tesla field is not sufficient for complete transformation. The reverse transformation occurs at lower field value resulting in a hysteretic field dependence of resistivity. At 50 K and 30K resistivity reaches its initial field value at 0 Tesla which indicates complete reverse transformation from FRI to AFM phase. Further cycling for these field values in negative direction results in mirror image of R-H curve obtained for positive field cycle. At lower temperatures anomalous behavior was observed as shown for 10 K and 5 K. With increasing magnetic field partial phase transformation from AFM to FRI phase is observed as 8 Tesla field is not sufficient to complete the transformation. With decreasing magnetic field reverse transformation starts at lower field with decreasing temperature and this transformation doesn't complete even when magnetic field is reduced to zero. Therefore zero field resistivity before and after the application of magnetic field shows large difference and this difference increases with decreasing temperature. For further cycling of magnetic field an envelope curve is obtained with much smaller variation of resistivity compared to virgin curve. Also the virgin curve lies outside the envelope curve. Earlier magnetization measurement on Co doped $Mn_2Sb$ also showed virgin curve lying outside the envelop curve at 4.2 K \cite{bart}. However in case of magnetization measurements, opening of hysteresis loop at zero field is not observed as magnetization goes to zero for the FRI state also. However the magnetic state at zero field is clearly demonstrated in R-H measurements in present study, where distinctly lower resistivity at zero field is observed after field cycling. 

The H-T diagram obtained from R-T and R-H measurements are shown in Figure 4. Transition temperature and field are taken as the minimum of the first order derivative of these curves, respectively. The transition temperature shifts to lower temperature with a rate 10K/Tesla for low field values which is consistent with earlier reports on Co doped $Mn_2Sb$ \cite{bart}.  The difference between the transition temperature during heating and cooling increases for higher field values $H\geq 5$ Tesla i.e. at low temperature. Such broadening of hysteretic region has been observed in many manganite system \cite{kuwa, toku}, transition metal doped FeRh \cite{bara1} and is generally associated with first order transition at low temperatures.  Besides broadening of hysteresis region this figure also shows the non-monotonic variation of lower critical field required for FRI to AFM transition. It has been shown in the case of $Nd_{0.5}Sr_{0.5}MnO_{3}$ \cite{rawa} that such non-monotonic behavior is anomalous and can not be explained in terms of conventional first order transition. Such behavior arises due to critically slow dynamics of the first order transition at low temperature and high temperature phase remains arrested at low temperatures. Therefore anomalous behavior in present case will also be discussed in terms of kinetic arrest and the decreasing lower critical field with decreasing temperature is a reflection of kinetic arrest band for this compound. The anomalous thermomagnetic irreversibility then can be explained due to arrested FRI phase at low temperatures.

It has been argued that field-induced first order magnetic transitions provide an experimentally versatile platform for studying glass-like kinetic arrest \cite{cha}. Since the transformation is H-induced, the transition temperature obtained under constant H cooling/heating must vary with H. For some systems this $T_C$ is low enough for kinetic arrest temperature $T_K$ to interfere and hinder the transition. Since $T_C$ varies with H one can vary H to go from a situation where $T_K < T^*$ to one where $T_K > T^*$ \cite{chad}. In the former case rapid cool-down is essential for glass-formation like in metglasses. In the latter case a comparatively slow cool-down can result in a glass, like in the glass-former O-terphenyl. Field induced first order magnetic transitions provide an experimental platform where the variation of H allows us to study very different glass-formation behavior in the same system; traversing novel paths in (H, T) space provide interesting new phenomena \cite{rawa, mane, bane, ban, sing, chat, chad, kuma, roy}. We now show that H-induced first order magnetic transitions also provide a fertile platform to study supercooled/superheated metastable states.

As has been argued earlier \cite{ban}, the slope of the $T_C(H)$ line is dictated by the magnetic order in the low-T zero field state and is negative in the present sample. This is shown in the schematic in figure 5(a) alongwith the supercooling and superheating spinodals $T^*(H)$ and $T^{**}(H)$. We now consider that we reach (in zero field) a certain temperature $T_0$ lying between $T^*(0)$ and $T_C(0)$. If we have reached $T_0$ by cooling, then the initial state is metastable supercooled FRI phase. A subsequent isothermal cycling of field at $T_0$ takes the sample from metastable FRI state to stable FRI state and back to metastable FRI state. If however, we reach $T_0$ by heating, then the starting H=0 state is stable AFM state. As H is raised, it transforms to stable FRI at $H^{**}(T_0)$ but returns to metastable FRI at H=0. The initial and final states at H=0 are thus different and we will observe an open loop in R vs H. This open loop will not be observed when $T_0$ is reduced to below $T^*$ or raised above $T^{**}$.

If the low-T zero field state is FRI, as shown in the schematic in figure 5(b), then the anomalous open loop will be observed on reaching $T_0$ by cooling in zero field (and not on heating) when $T_0$ will be between $T^*(0)$ and $T^{**}(0)$. This experimental check for an open loop on cooling vs. heating thus cross-checks the low-T-zero-H state. The open loop will not be observed as $T_0$ is lowered below $T^*$ in this case also. This latter feature contrasts with the case of kinetic arrest where the open loop becomes more prominent as $T_0$ is lowered \cite{rawa, kuwa, mane}.

Figure 6 shows the R-H curve for temperatures 60 K, 120 K, 130 K and 160K. For each of these measurements sample was cooled down to 5K and then warmed to the measurement temperature under zero field condition before taking measurements. Since 60 K and 160 K are below $T^*(0)$ and above $T^{**}(0)$ respectively; the initial ZFCW state at H=0 are stable states as also are the remnant states. At 60 K we observe a field induced transition but not at 160 K. At 120 K and 130 K the initial ZFCW states are AFM (with higher resistance) but the remnant  states are FRI (with lower resistance). This is consistent with the schematic shown in figure 5 (a). Because of disorder-broadening, the initial state will also have a stable FRI phase coexisting along with AFM phase. With the application of magnetic field the AFM phase (both stable and metastable) will transform to FRI phase completely on crossing the $H^{**}$ band. However with reducing magnetic field  it will not transform back to initial state and reverse transformation will take place only for those regions of sample for which $T^*$ line is crossed on reducing magnetic field. Since the state of the system is different before and after the application of magnetic field, we observe an open hysteresis loop at these temperatures. Further field cycling produces symmetric envelope curve. This envelope curve will be similar to R-H curve at the same temperature when reached during cooling . This is shown in figure 6 for 120 K and 130 K, where R-H measurement were performed at these temperatures after cooling sample from room temperature under zero field condition. The R-H curve during cooling shows no difference between virgin curve and envelope curve at these temperatures and these R-H curves overlaps with the respective R-H  envelope curve obtained during heating.

We have shown the distinct behavior of metastable supercooled state from the metastable kinetically arrested state at low temperature. This brings out very clearly the difference between supercooling and glass-like kinetic arrest. Now we study the transformation of kinetically arrested FRI state to stable AFM state. For this we controlled the arrested FRI fraction by field annealing and its transformation is studied as function of magnetic field and temperature. For these measurements sample is cooled from room temperature to 5 K under an applied magnetic field $H_{an}$. Then at 5 K, magnetic field is reduced isothermally to 0 Tesla and resistivity is measured as a function of magnetic field which is shown in Fig 7. The lower resistance at higher $H_{an}$ indicates larger fraction of FRI phase, and brings out the tunability of phase fractions at 5 K. As can be seen in Fig 7, R-H curve remains constant initially and then shows a rapid increase with decreasing magnetic field indicating the de-arrest of FRI phase in to AFM phase. R-H curve for $H_{an}\geq 3-8$ Tesla annealing field merges with lowering magnetic field whereas the 1 Tesla curve remains at higher values. In fact resistivity for 1 Tesla annealing field remains almost constant and is distinctly higher compared to that for higher annealing fields. This implies that $H_1$ (figure 1(b))for this sample could be lying above 1 Tesla magnetic field. These trends are similar to that expected for anticorrelated supercooling and kinetic arrest bands i.e. region which have lower supercooling temperature are arrested at higher temperature \cite{kuma, cha, roy}.
	
	The kinetically arrested FRI phase can show de-arrest with increasing temperature also. In such case FRI phase transform to AFM phase with increasing temperature. To demonstrate this, sample is cooled in the presence of $H_{an}$ to 5 K and then magnetic field is changed to 4 Tesla isothermally. Then R vs. T measurement are carried out during heating in the presence of 4 Tesla, results of which are shown in figure 8. A reentrant FRI to AFM to FRI transitions is clearly visible for all the FCW curves for which $H_{an}$ is larger than the measurement field of 4 Tesla. Whereas, only one AFM to FRI transition is observed for $H_{an}\leq 4.0$ Tesla. This is brought out very clearly in the inset where we plot temperature derivative of resistivity ($dR/dT$) for $H_{an}$ = 2 tesla and 4.5 Tesla. The fact that two transitions are seen only when annealing field $H_{an}$ is larger than measuring field is consistent with the low temperature low field ground state being AFM \cite{ban}. The AFM to FRI transition is independent of annealing field  $H_{an}$ whereas the FRI to AFM transition depends on $H_{an}$. For lower $H_{an}$, FRI to AFM transition starts at higher temperature compared to higher $H_{an}$. It is evident from this figure resistivity for $H_{an}\leq 5.5$ Tesla remains constant initially and then shows an upturn with increasing temperature. The temperature where resistivity shows an upturn is higher for lower $H_{an}$. Also the curves for higher $H_{an}$ merge at lower temperature before merging to lower $H_{an}$. And all the curves for $H_{an}\geq 4.5$ Tesla merges before merging to FC 4 Tesla curve during warming. All these  features are similar to those seen in earlier studies on very different materials and the de-arrest as function of temperature once again is similar to that expected for anti-correlated supercooling and kinetic arrest band \cite{kuma, cha} . 

\maketitle\section{Conclusions}
To conclude a detailed investigation of first order FRI to AFM transition is carried out at low temperature. These studies reveal anomalous thermomagnetic irreversibility along with non-monotonic variation of lower  critical field at low temperatures. These anomalies are interpreted in terms of kinetic arrest of FRI phase due to critically slow dynamics of transformation on the measurement time scale. The irreversibility due to kinetic arrest are distinguished from those arising due to metastability in the supercooling or superheating regime. Below transition temperature both FRI and AFM phase can coexist over a wide temperature and magnetic field range and tunability of these phases is demonstrated. The measurement also show anticorrelation between kinetic arrest band and supercooling band. Similar anticorrelation has been observed earlier for AFM to FM transition in intermetallic and manganite systems. This study extends this universality to FRI to AFM transition. The present compound, due to its less complex structure and interactions, may work as model system for manganites showing similar behavior.

\maketitle\section{Acknowledgments}

We acknowledge Dr. Alok Banerjee for valuable discussion and suggestions.

\maketitle\section{Figure Captions}

Figure 1: (a) Resistivity as a function of temperature in the absence of magnetic field during cooling and heating cycle. Inset shows the derivative of resistivity around transition temperature. (b) Schematic of kinetic arrest band and supercooling band in H-T space adapted from Banerjee et al. \cite{ban}.The co-existing phases will be observed only if cooling or annealing field H lies between $H_1$ and $H_2$. 

Figure 2: Temperature dependence of resistivity in the presence of  various constant magnetic field. The measurements were taken in the presence of labeled magnetic field where ZFCW is measured during warming after zero field cooling, FCC during cooling in field and FCW during warming after field cooling in same field.  Scale on Y-axis is for 0 Tesla curve and other curves are shifted downward for the sake of clarity. The resistivity values at 300 K are independent of H.

Figure 3: Isothermal R vs H for $Mn_{1.85}Co_{0.15}Sb$ at various temperatures. The open R-H loop and the virgin curve lying outside envelope curve are highlighted at low temperature. 

Figure 4: H-T phase diagram for $Mn_{1.85}Co_{0.15}Sb$ obtained from R-H (crossed squares) and R-T measurements (solid circles). See text for details.

Figure 5: Transformation in the supercooling or superheating regime for different paths are demonstrated along with corresponding schematic of free energy diagram. The state in which the system exists is indicated by the filled circle. Note that in the case (a) an open hysteresis loop will be seen only if $T_0$ is reached on heating; and in case (b) only if $T_0$ is reached on cooling. In these cases the virgin state and remnant states are different; and the remnant state is independent of whether $T_0$ is reached by heating or cooling.
 
Figure 6: Magnetic field dependence of resistivity (R-H) at $T_0$ = 60 K, 160 K, 120 K and 130 K with $T_0$ being reached by heating from 5 K. The different zero field resistivity before and after the application of magnetic field during heating at 120 K (c) and 130 K (e) is highlighted, whereas no such difference is observed when $T_0$  = 120 K (d) and 130 K (f) is reached by cooling, consistent with figure 5 (a).

Figure 7: De-arrest of FRI phase is shown with reducing magnetic field at 5 K in R-H measurement for samples cooled under various annealing field $H_{an}$. The merger of R-H curve with reducing magnetic field suggest anticorrelation \cite{cha}. See text for details.

Figure 8: Reentrant phase transition from FRI to AFM to FRI state is shown in the temperature dependence of resistivity in the field of 4 Tesla during warming. Samples were cooled under 0, 2.0, 4.0, 4.5, 5.0, 5.5 and 6.0 Tesla annealing field $H_{an}$. The inset highlights that two transitions are seen only when cooling (or annealing) field is larger than measuring field which is consistent with the low-T low-H ground state being AFM \cite{ban}. Curve for higher $H_{an}$ merges at lower temperature showing anticorrelation between the supercooling and kinetic arrest band \cite{cha}. See text for details.

\begin{figure*}[t]
	\centering
	\includegraphics[width=9cm]{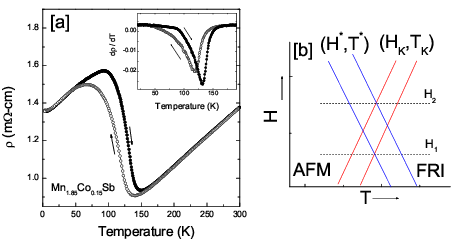}
  \caption{(a) Resistivity as a function of temperature in the absence of magnetic field during cooling and heating cycle. Inset shows the derivative of resistivity around transition temperature. (b) Schematic of kinetic arrest band and supercooling band in H-T space adapted from Banerjee et al. \cite{ban}.The co-existing phases will be observed only if cooling or annealing field H lies between $H_1$ and $H_2$.}
	\label{fig:Graph1}
\end{figure*}

\begin{figure*}[t]
	\centering
		\includegraphics[width=7cm]{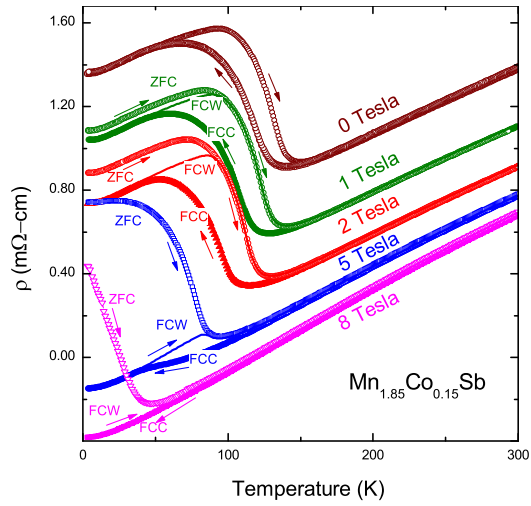}
  	\caption{Temperature dependence of resistivity in the presence of  various constant magnetic field. The measurements were taken in the presence of labeled magnetic field where ZFCW is measured during warming after zero field cooling, FCC during cooling in field and FCW during warming after field cooling in same field.  Scale on Y-axis is for 0 Tesla curve and other curves are shifted downward for the sake of clarity. The resistivity values at 300 K are independent of H.}
	\label{fig:Graph2}
\end{figure*}

\begin{figure*}[t]
	\centering
		\includegraphics[width=8.5cm]{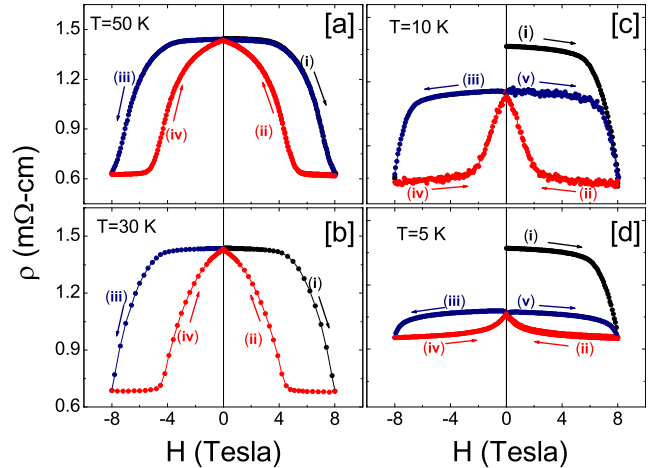}
		\caption{Isothermal R vs H for $Mn_{1.85}Co_{0.15}Sb$ at various temperatures. The open R-H loop and the virgin curve lying outside envelope curve are highlighted at low temperature. }
\end{figure*}

\begin{figure*}[t]
	\centering
		\includegraphics[width=7cm]{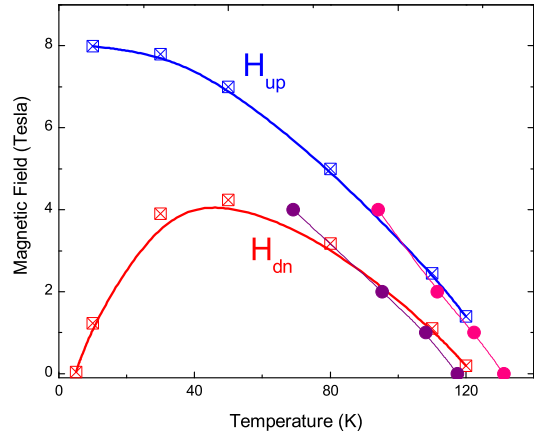}
		\caption{H-T phase diagram for $Mn_{1.85}Co_{0.15}Sb$ obtained from R-H (crossed squares) and R-T measurements (solid circles). See text for details.}
\end{figure*}

\begin{figure*}[t]
	\centering
		\includegraphics[width=14cm]{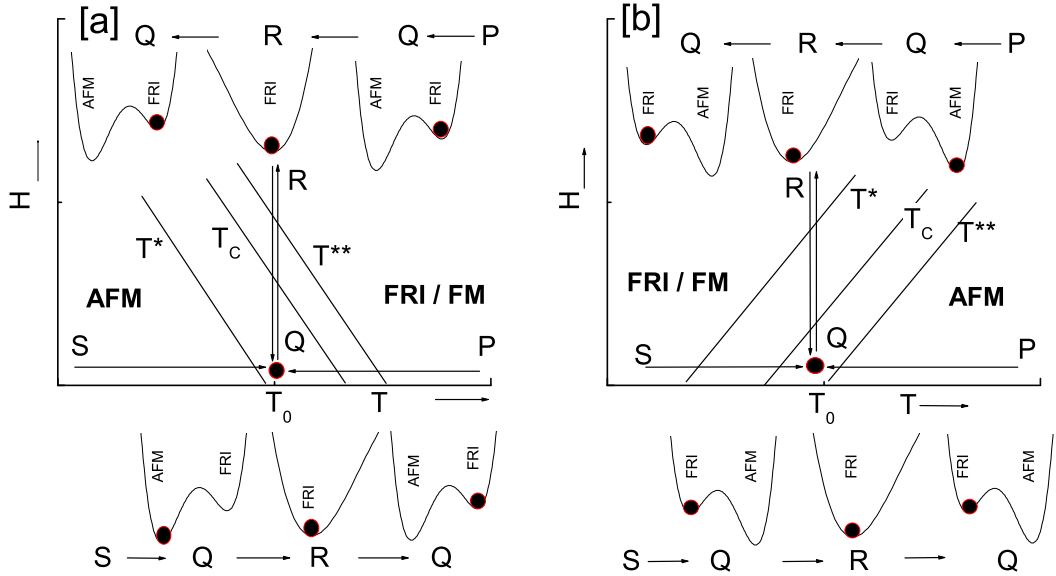}
		\caption{Transformation in the supercooling or superheating regime for different paths are demonstrated along with corresponding schematic of free energy diagram. The state in which the system exists is indicated by the filled circle. Note that in the case (a) an open hysteresis loop will be seen only if $T_0$ is reached on heating; and in case (b) only if $T_0$ is reached on cooling. In these cases the virgin state and remnant states are different; and the remnant state is independent of whether $T_0$ is reached by heating or cooling}
\end{figure*}

\begin{figure*}[t]
	\centering
		\includegraphics[width=14cm]{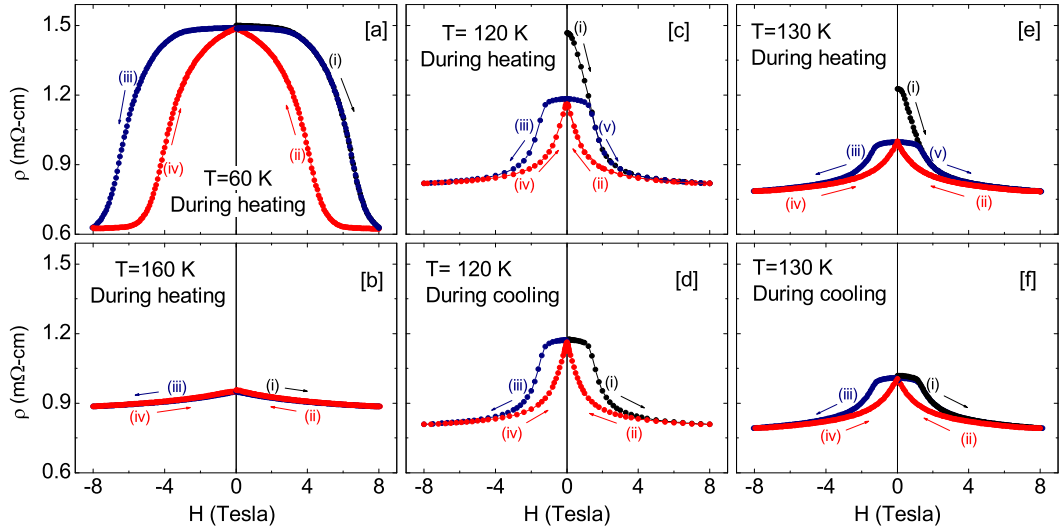}
		\caption{Magnetic field dependence of resistivity (R-H) at $T_0$ = 60 K, 160 K, 120 K and 130 K with $T_0$ being reached by heating from 5 K. The different zero field resistivity before and after the application of magnetic field during heating at 120 K (c) and 130 K (e) is highlighted, whereas no such difference is observed when $T_0$  = 120 K (d) and 130 K (f) is reached by cooling, consistent with figure 5 (a).}
\end{figure*}

\begin{figure*}[t]
	\centering
		\includegraphics[width=7cm]{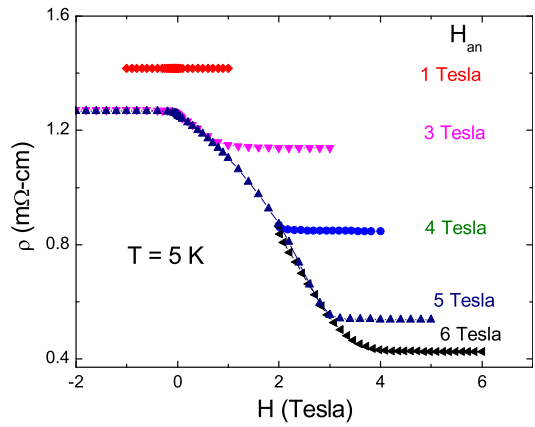}
		\caption{De-arrest of FRI phase is shown with reducing magnetic field at 5 K in R-H measurement for samples cooled under various annealing field $H_{an}$. The merger of R-H curve with reducing magnetic field suggest anticorrelation \cite{cha}. See text for details.}
\end{figure*}

\begin{figure*}[t]
	\centering
		\includegraphics[width=7cm]{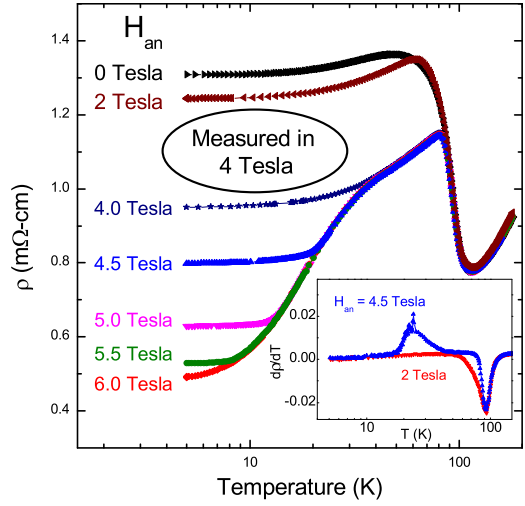}
		\caption{Reentrant phase transition from FRI to AFM to FRI state is shown in the temperature dependence of resistivity in the field of 4 Tesla during warming. Samples were cooled under 0, 2.0, 4.0, 4.5, 5.0, 5.5 and 6.0 Tesla annealing field $H_{an}$. The inset highlights that two transitions are seen only when cooling (or annealing) field is larger than measuring field which is consistent with the low-T low-H ground state being AFM \cite{ban}. Curve for higher $H_{an}$ merges at lower temperature showing anticorrelation between the supercooling and kinetic arrest band \cite{cha}. See text for details.}
\end{figure*}

\end{document}